\documentclass{article}

\usepackage{arxiv}

\usepackage[utf8]{inputenc}
\usepackage[T1]{fontenc}
\usepackage[hidelinks]{hyperref}
\usepackage{url}
\usepackage{booktabs}
\usepackage{amsthm}
\usepackage{amsmath}
\usepackage{amssymb}
\usepackage{amsfonts}
\usepackage{siunitx}
\usepackage{tikz}
\usepackage{pgfplots}
\pgfplotsset{compat=1.18}
\pgfplotsset{every axis legend/.append style={font=\tiny}}
\usepackage{subcaption}
\usepgfplotslibrary{groupplots}
\usepackage{longtable}
\usepackage{array}
\usepackage{multirow}
\usepackage{tabularx}
\usepackage{ragged2e}
\newcommand{\RR}{\RaggedRight\arraybackslash}
\usepackage[title]{appendix}
\usepackage{xcolor}
\usepackage{enumitem}

\definecolor{ruleblack}{RGB}{30,30,30}
\definecolor{lightgray}{RGB}{120,120,120}
\definecolor{flagred}{RGB}{190,40,30}

\usepackage[labelfont=bf,font=small,labelsep=colon]{caption}

\title{Mission-Centric Requirements Analysis of Model Predictive Control\\
in Spacecraft Rendezvous and Proximity Operations}

\author{
    Nils Maier \\
    University of Stuttgart \\
    \texttt{st172065@stud.uni-stuttgart.de} \\
    \AND
    Torbjørn Cunis \\
    Institute of Flight Mechanics and Controls \\
    University of Stuttgart \\
    \texttt{torbjoern.cunis@ifr.uni-stuttgart.de} \\
    \And
    Tam W.\ Nguyen \\
    Department of Electrical Engineering \\
    Kyoto University \\
    \texttt{nguyen.tamwilly.3e@kyoto-u.ac.jp} \\
}

\renewcommand{\shorttitle}{Mission-Centric MPC for Spacecraft RPO}
\renewcommand{\headeright}{Technical Report}
\renewcommand{\undertitle}{Technical Report --- ADRIOS / ClearSpace-1 Application}

\hypersetup{
pdftitle={Mission-Centric Requirements Analysis of Model Predictive Control in Spacecraft Rendezvous and Proximity Operations},
pdfsubject={eess.SY, math.OC},
pdfauthor={Nils Maier},
pdfkeywords={Economic Model Predictive Control, Spacecraft Rendezvous, Proximity Operations, Active Debris Removal, Dissipativity},
}

\begin{document}
\maketitle

\begin{center}
\begin{minipage}{0.92\textwidth}
\small\itshape
\textbf{Publication Notice.}
This report documents a collaboration between the authors in the scope of a Master's thesis research exchange between the University of Stuttgart and Kyoto University (GESR programme). The report may unilaterally be published
on arXiv, made public on a university's repository, and/or submitted for
publication in a scientific journal by any author or subset of authors,
contingent on the following conditions:

\begin{enumerate}[leftmargin=2em, topsep=0.4em, itemsep=0.2em]
  \item The list and order of authors remain unchanged and the report is
        published, made public, or submitted for publication in the present
        form.
  \item The publication is not effective before December~31, 2026.
  \item The content of the report has not yet been published, made public,
        or submitted for publication in full or in parts.
  \item The authors have not agreed otherwise in writing.
\end{enumerate}

This agreement does not prevent the authors from unanimously publishing,
making public, or submitting for publication the content of this report,
in full or in parts, at any time.

\medskip
\noindent The Authors \hfill Stuttgart\,/\,Kyoto, July~2026
\end{minipage}
\end{center}

\bigskip
\hrule
\bigskip
\newpage

\begin{abstract}
Rendezvous and proximity operations (RPO) missions are commonly designed as
a sequence of distinct phases each governed by its own sample time, constraint set, and cost
structure, with the translation from mission-level requirements to
model predictive control (MPC) design parameters typically performed ad hoc.
This report takes a first step toward a systematic, traceable mapping
between mission requirements and MPC design choices, using the ADRIOS /
ClearSpace-1 active debris removal mission as a case study across three
sequential operations. Its central hypothesis is that Economic MPC (EMPC),
in which the stage cost reflects propellant consumption directly, is
structurally better suited to proximity operations than fixed-point
regulation or reference tracking, because it can exploit the natural,
fuel-free relative-orbital-motion trajectories admitted by the
Clohessy--Wiltshire--Hill (CWH) dynamics instead of fighting them. We show
that a pure-fuel stage cost is fundamentally degenerate for this class of
problem as the closed-loop optimum is not strictly dissipative with respect
to any single target state or orbit, and analyse, operation by operation,
which terminal ingredient (a terminal region, a terminal cost, a terminal
equality, or a stage-cost regularisation) is required to repair this
degeneracy, at what fuel and time cost, and under what formal guarantee.
Across all three operations the economic controllers cut fuel consumption
by roughly 40--99\% relative to standard quadratic-tracking MPC, at the
cost of longer manoeuvre times, and we synthesise the results into a
cross-scenario table of mission requirements, MPC ingredients, and their
residual limitations.
\end{abstract}


\newpage
\section*{List of Acronyms}
\addcontentsline{toc}{section}{List of Acronyms}
\begingroup
\renewcommand{\arraystretch}{1.3}
\begin{longtable}{@{}p{2.2cm}p{11cm}@{}}
ADR    & Active Debris Removal \\
CAM    & Collision Avoidance Manoeuvre \\
CLF    & Control Lyapunov Function \\
ConOps & Concept of Operations \\
CW     & Clohessy--Wiltshire \\
CWH    & Clohessy--Wiltshire--Hill \\
DARE   & Discrete Algebraic Riccati Equation \\
ECI    & Earth-Centred Inertial \\
EMPC   & Economic Model Predictive Control \\
ESA    & European Space Agency \\
FAP    & Final Approach Point \\
FKP    & Formation Keeping Point \\
GNC    & Guidance, Navigation and Control \\
IPP    & Initial Proximity Point \\
KOS    & Keep-Out Sphere \\
LEO    & Low Earth Orbit \\
LEOP   & Launch and Early Orbit Phase \\
LQR    & Linear Quadratic Regulator \\
LTI    & Linear Time-Invariant \\
MPC    & Model Predictive Control \\
NMC    & Natural Motion Circumnavigation \\
QP     & Quadratic Program(ming) \\
RPO    & Rendezvous and Proximity Operations \\
ZOH    & Zero-Order Hold \\
ISS    & International Space Station \\
SRP    & Solar Radiation Pressure \\
\end{longtable}
\endgroup

\newpage
\section*{Nomenclature}
\addcontentsline{toc}{section}{Nomenclature}
\begingroup
\renewcommand{\arraystretch}{1.3}
\begin{longtable}{@{}p{1.8cm}p{9.5cm}p{2cm}@{}}
\toprule
Symbol & Description & Unit \\
\midrule
\endhead
$n$                     & Target mean motion & rad/s \\
$T$                     & Orbital period & s \\
$\Delta t$              & Sample time & s \\
$N$                     & Prediction horizon length & steps \\
$x,y,z$                 & Hill-frame relative position (radial, along-track, cross-track) & m \\
$\dot x,\dot y,\dot z$  & Hill-frame relative velocity components & m/s \\
$\boldsymbol{x}$        & State vector $[x,y,z,\dot x,\dot y,\dot z]^\top$ & --- \\
$u$                     & Control input (specific thrust) & m/s$^2$ \\
$u_{\max}$              & Maximum thrust / acceleration bound & m/s$^2$ \\
$A_c,\,B_c$             & Continuous-time system matrices & --- \\
$A_d,\,B_d$             & Discrete-time (ZOH) system matrices & --- \\
$Q,\,R$                 & Stage-cost weighting matrices (state, input) & --- \\
$Q_{\mathrm{dare}},\,R_{\mathrm{dare}}$ & DARE design weights & --- \\
$P$                     & Terminal cost matrix (DARE solution) & --- \\
$K$                     & Stabilising feedback gain & --- \\
$\ell$                  & Stage cost function & --- \\
$\lambda$               & Storage function (dissipativity) & --- \\
$\rho(\cdot)$           & Class-$\mathcal{K}_\infty$ function (dissipativity margin) & --- \\
$\rho_c$                & Stage-cost regularisation weight & --- \\
$\varepsilon$           & Regularisation weight, augmented stage cost & --- \\
$x_s,\,u_s$             & Optimal economic steady state & --- \\
$x_{\mathrm{ref}}$      & Generic reference/target state & --- \\
$x_{\mathrm{dock}}$     & Docking-port target state (Op.\ 3) & --- \\
$x_{\mathrm{switch}}$   & Handover-point target state (Op.\ 1) & --- \\
$e$                     & Error state, $e = x - x_{\mathrm{ref}}$ & --- \\
$\Delta v$              & Total characteristic velocity (fuel usage) & m/s \\
$r_{\mathrm{KOS}}$      & Keep-out sphere radius & m \\
$\phi_0$                & Initial phase angle of NMC orbit & rad \\
$y_c$                   & Along-track drift centre of NMC orbit & m \\
$b$                     & NMC ellipse radial semi-axis (used interchangeably with $\rho$ in Op.\ 2/3 discussion); along-track semi-axis is $2b$ & m \\
\bottomrule
\end{longtable}
\endgroup

\newpage

\section{Introduction}
\label{sec:intro}

Active debris removal (ADR) missions are structured as a sequence of distinct proximity operations. A far-range closing approach, an inspection or fly-around phase, and a final docking or capture manoeuvre \cite{Fehse_2003,Petersen2023}. Model Predictive Control (MPC) is well suited to this kind of mission because it handles input and state constraints explicitly over a receding horizon \cite{MAYNE2000789,Rawlings2017}, and because the same underlying optimisation problem can, in principle, be re-parameterised from one operation to the next rather than re-designed from scratch. MPC has accordingly become a leading approach for constrained spacecraft rendezvous and proximity operations \cite{Hartley2015,Weiss2015,DiCairano2012,DiCairano2018}.

In practice, however, the translation from mission-level requirements like orbital regime, approach-corridor geometry, fuel budget and keep-out zones to MPC design parameters like horizon length, cost structure, terminal ingredients and constraint formulation is typically performed ad hoc, on a per-mission and per-phase basis, without a systematic methodology connecting the two.

This is the gap the thesis tries to close. The eventual goal is a requirements framework that maps mission parameters onto MPC design choices in a traceable, reusable way, developed across an entire mission sequence rather than a single isolated manoeuvre.

The present report is a first step towards it. Its central working hypothesis is that Economic MPC (EMPC) \cite{Diehl2011,Angeli2012,Ellis2014,Faulwasser2018}, in which the stage cost reflects the physical resource of interest directly, e.g. propellant, is structurally better suited to proximity operations than fixed-point regulation or reference-tracking MPC, because it can exploit natural relative-orbital-motion dynamics instead of fighting them.

The ADRIOS / ClearSpace-1 mission \cite{Woicke2025} is used here to derive the framework. The mission serves two purposes:
\begin{enumerate}
    \item Expose which problems arise when an MPC is constructed for a specific proximity operation
    \item Identify where EMPC is useful, why it is useful and what mission requirements creates that need
\end{enumerate}
 This report documents the current implementation status of this first application to the ADRIOS / ClearSpace-1 Concept of Operations (ConOps) \cite{Woicke2025}, treated as three sequential operations.
 \begin{enumerate}
     \item Operation 1 : far-range closing
     \item Operation 2 : fly-around by using the Natural Motion Circumnavigation arising from the CW-equations \cite{schaub2018analytical}
     \item Operation 3 : final docking
 \end{enumerate}
Section~\ref{sec:mission} lays out the above sequence and the requirements each phase imposes; Section~\ref{sec:foundations} establishes the rendezvous dynamics and how they have nice properties for one specific Rendezvous and Proximity Operation (RPO) mission; Section~\ref{sec:whyempc} argues why MPC and especially EMPC is the right tool for these dynamics; Section~\ref{sec:mpc} introduces the shared EMPC formulation and design axis used across all three operations; Sections~\ref{sec:op1}--\ref{sec:op3} report the implementation and results for each operation in turn; Section~\ref{sec:cross} draws the results together into a cross-scenario requirements comparison; and Section~\ref{sec:conclusion} summarises what is established and what remains open.

\section{Mission Overview \& Scenarios}
\label{sec:mission}

\subsection{ADRIOS ClearSpace-1}

ADRIOS ClearSpace-1 is the first European mission to demonstrate in-orbit rendezvous, capture, and removal of an existing non-cooperative space object from low Earth orbit \cite{Woicke2025}. The selected client is PROBA-1, an ESA microsatellite (launched 2001, approx.\,94\,kg, $60 \times 60 \times 80$\,cm) in a Sun-synchronous orbit at roughly \SI{400}{km} altitude. The mission is developed by OHB System AG and ClearSpace SA under ESA's Space Safety Programme, with launch expected in 2028.

\subsection{Concept of Operations}
\label{subsec:conops}

After launch, parallel commissioning and phasing bring the servicer to a switch point approximately \SI{30}{km} behind PROBA-1. At this point absolute navigation hands over to relative GNC, initiating the rendezvous and proximity operations (RPO) phase. This far-to-close-range phasing follows the classical rendezvous sequence \cite{Fehse_2003,Woffinden2017}. The full ADRIOS ConOps comprises the following sequential phases \cite{Woicke2025}:

\begin{enumerate}
  \item \textbf{Closing.} Far-range transfer from \,30\,km to a few hundred metres, using passively safe impulsive manoeuvres and angles-only optical navigation. A ranging device supplements the camera in the final portion.
  \item \textbf{Fly-around.} The servicer enters NMC orbits around the client at 50--100\,m, observing and characterising PROBA-1 with mid-range navigation. The same guidance concept as closing is used; range measurements augment the camera for improved robustness.
  \item \textbf{Proximity manoeuvring.} The servicer transfers to the Formation Keeping Point (FKP) at \,150\,m along-track, hops to the Initial Proximity Point (IPP), and then advances to the Final Approach Point (FAP) using close-range navigation and motion synchronisation to track PROBA-1's rotation. A rehearsal of the full sequence precedes each capture attempt; the mission is sized for three capture attempts with one Collision Avoidance Manoeuvre (CAM) budgeted per attempt.
  \item \textbf{Capture.} Once PROBA-1 is inside the capture volume, GNC hands off to the capture system, which closes four robotic arms around the client. After securing the client, the stack detumbles and a perigee-lowering manoeuvre initiates deorbit within five years.
\end{enumerate}

\subsection{Scope of This Report}
This report addresses the RPO GNC problem for a simplified version of the first three phases of the above sequence. The proximity manoeuvring phase (FKP $\to$ FAP and motion synchronisation) requires six-degrees-of-freedom pose-tracking control and is outside the scope of the present work. Thus, Operation 3 deals with a simplified terminal approach, which takes the chaser from fly-around to the targets eventual docking port, where the capture system takes over. The three operations studied here are treated not as independent test cases but as one continuous mission sequence in which the same design translation (mission requirement to MPC ingredients) is instantiated three times under changing range and dynamics regime. Table~\ref{tab:conops} lists the sequence. Each row corresponds to a single design instance, in which the sample time, horizon and terminal ingredient are selected according to that phase's requirements. $\Delta t$ is here the ZOH discretisation step of the prediction model and the rate at which a new control input is
computed and applied.\\
In Operation~1, $\Delta t = \SI{120}{s}$ is loose relative to the CW
timescale ($T=\SI{5545}{s}$) because far-range closing involves slowly
evolving relative motion and a wide (\SI{50}{m}) convergence tolerance, so
finer sampling would add computational cost without improving control
authority. In Operation~2, $\Delta t = T/92 \approx \SI{60.27}{s}$ is fixed by the requirement that the horizon land
exactly on one full orbital period, $N\Delta t = T$, since the periodic
terminal equality~\eqref{eq:nmc-termeq} needs the predicted trajectory to
close the NMC orbit exactly rather than approximately. In Operation~3,
$\Delta t = \SI{10}{s}$ is an order of magnitude finer because the
docking-approach line-of-sight cone and the sub-metre terminal tolerance
both tighten as range decreases, and a coarser sample would let the chaser
drift outside the corridor between updates. The horizon length $N$ is
likewise set by a different requirement in each case (see
Section~\ref{sec:operations})

\begin{table}[h]
\centering
\caption{Operations covered in this report (from the ADRIOS ConOps), and the mission requirement that drives each MPC design choice.}
\label{tab:conops}
\begin{tabularx}{\textwidth}{@{}c >{\RR}X c c c >{\RR}X@{}}
\toprule
Op & Scenario / Mission requirement & Range & $\Delta t$ / $N$ & Horizon ($N\Delta t$) & MPC design choice \\
\midrule
1 & Far-range closing / reach box around \SI{300}{m} switch point & 10\,km $\to$ 300\,m & 120\,s / 20 & \SI{2400}{s} ($\approx$\SI{40}{min}) & Pure fuel + $\ell_\infty$ terminal ball; $+\,\rho_c\|e\|^2$ regularisation; $+$ CLF terminal cost ($M_1\succeq0$) \\
2 & NMC fly-around / acquire NMC orbit & 300\,m $\to$ NMC orbit & $\approx$60.27\,s / 92 & \SI{5545}{s} ($=T\approx\SI{1.54}{h}$) & Pure fuel + periodic terminal eq.\ $x_N=x^*_{\sigma(k)}$; $+\,\varepsilon\|x-x^*_k\|^2$ phase-sync \\
3 & Final docking / reach docking-port vicinity & 150\,m $\to$ $<$1\,m & 10\,s / 24 & \SI{240}{s} ($=\SI{4}{min}$) & Pure fuel + CLF terminal cost; $+$ hard terminal eq.\ $x_N=x_{\mathrm{dock}}$ \\
\bottomrule
\end{tabularx}
\end{table}

\section{Rendezvous Dynamics}
\label{sec:foundations}

\subsection{Rendezvous Mechanics and the CW Equations}
\label{sec:cw}

Spacecraft rendezvous requires driving the relative state of a chaser vehicle to a desired position/velocity relative to a target. The full two-body problem in an Earth-Centred Inertial (ECI) frame is nonlinear. For small relative separations and near-circular reference orbits, the relative dynamics linearise to the Clohessy--Wiltshire (CW) equations \cite{Hill1878,ClohessyWiltshire1960,schaub2018analytical}:

\begin{align}
\ddot{x} - 2n\dot{y} - 3n^2 x &= u_x, \\
\ddot{y} + 2n\dot{x} &= u_y, \\
\ddot{z} + n^2 z &= u_z,
\end{align}
with state vector $[x,\,y,\,z,\,\dot{x},\,\dot{y},\,\dot{z}]^\top$. These form a 6-state, 3-input linear time-invariant (LTI) system.

The CW equations rest on a circular target orbit (constant $n$), unperturbed two-body gravity (no $J_2$, drag, or SRP), and a first-order linearisation of the differential gravity field valid only for relative separations small compared to the orbital radius \cite{Fehse_2003,schaub2018analytical}. It further assumes the chaser's mass and thrust are too small to perturb the target's trajectory, and that the Hill frame rotates at the constant rate $n$. All three operations in this report satisfy these conditions comfortably so the CW model is expected to hold with negligible mismatch error throughout.

A key structural property is visible directly in the equations as $x$ and $y$ are coupled through the Coriolis terms ($2n\dot{y}$, $2n\dot{x}$), while $z$ decouples entirely and behaves as a simple harmonic oscillator at frequency $n$. The in-plane ($x$-$y$) motion combines a bounded oscillatory mode with a secular along-track drift term that grows linearly in time. The CW system is therefore not asymptotically stable. Only under a specific initial-condition constraint, the drift-free condition introduced in Section~\ref{sec:nmc}, does the secular term vanish, leaving a closed, periodic orbit (Natural Motion Circumnavigation, NMC). More generally, the unforced ($u \equiv 0$) response is dominated by the along-track drift \cite{schaub2018analytical,Fehse_2003}. This structural instability is the fundamental challenge for rendezvous control but it is also, for EMPC specifically, a resource to be exploited.

\subsection{Natural Motion Circumnavigation}
\label{sec:nmc}
Among the unforced solutions of the CW equations (Section~\ref{sec:cw}), one
family is of particular operational value. The closed, periodic relative orbits
known as Natural Motion Circumnavigations (NMC). Because they require no
propellant to maintain, they are the natural target regime for the fly-around
inspection phase \cite{schaub2018analytical,Fehse_2003}.
\begin{figure}[ht]
    \centering
    \resizebox{0.9\textwidth}{!}{\input{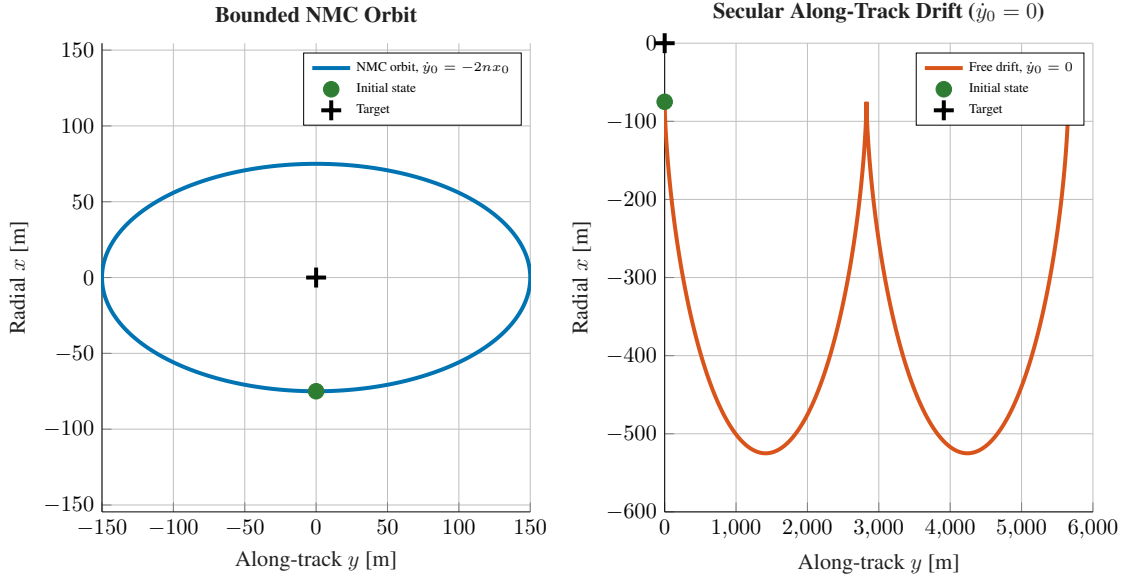}}
    \caption{Bounded $2{:}1$ NMC ellipse in the Hill frame under the drift-free
             condition $\dot{y}_0 = -2n\,x_0$. The chaser traces a closed,
             periodic orbit about the target with no along-track drift. When
             this condition is violated (e.g.\ $\dot{y}_0 = 0$), the
             uncancelled secular term causes the along-track coordinate $y$ to
             grow linearly in time, and the trajectory drifts away from the
             target instead of closing.}
    \label{fig:nmc_ellipse}
\end{figure}
The along-track solution of the CW equations carries a secular term $\left(6n\,x_0 + 3\dot{y}_0\right)$ that grows linearly in time and drives the
chaser away from the target. Bounded motion therefore requires this term to
vanish, which gives the drift-free condition
\begin{equation}
    \dot{y}_0 = -2n\,x_0 .
    \label{eq:drift_free}
\end{equation}
When \eqref{eq:drift_free} holds, the chaser executes a closed, periodic orbit
relative to the target. When it is violated (e.g.\ $\dot{y}_0 = 0$), a secular
along-track drift accumulates, since the $y$-dynamics carry no restoring force.

With the drift-free condition and $\dot{x}_0 = 0$, the in-plane motion reduces to
\begin{align}
    x(t) &= x_0 \cos nt, \\
    y(t) &= -2 x_0 \sin nt + y_0 ,
\end{align}
a closed ellipse with radial semi-axis $|x_0|$ and along-track semi-axis
$2|x_0|$. The geometry acquired in Operation~2 (Section~\ref{sec:op2}).

\section{Motivation: Why (Economic) Model Predictive Control?}
\label{sec:whyempc}

Classical regulation techniques, like unconstrained PD control or LQR do not represent mission constraints explicitly. Rendezvous and Proximity Operations (RPO) is therefore a poor match for these techniques, because nearly every requirement a proximity operation has to satisfy is some kind of constraint, e.g. keep-out zones and passive-safety envelope constrain the state, thrust and total-$\Delta v$ budgets bound the input and line-of-sight cones bound the geometry between chaser and target \cite{Petersen2023,Hartley2015}. This is where MPC yields advantages over classical regulation techniques. It poses the control problem as a constrained, finite-horizon optimisation, re-solved on a receding horizon. This allows the inclusion of state and input constraints in the problem formulation \cite{MAYNE2000789,Rawlings2017}. Therefore MPC is the natural candidate for proximity operations as you can also see in the extensive literature published in the last years \cite{Hartley2015,Weiss2015,DiCairano2012,DiCairano2018}.

A second reason for using MPC in RPO specifically is the sequential structure of the RPO missions. Throughout their phases, far-range closing to docking, the sample time, active constraint set, and stage/terminal cost weighting all change from phase to phase, driven by the tightening safety margins and update-rate requirements as range decreases. What remains fixed across this entire sequence is the underlying relative-motion model (the CWH dynamics themselves), the condensed QP structure, and the solver architecture used to solve it \cite{Fehse_2003,Petersen2023}. Weiss et al.\ demonstrated exactly this separation. A rendezvous-phase QP with a long sampling period is switched to a docking-phase QP with a shorter sampling period and a different constraint set, while the prediction model and QP solver architecture are carried across the switch unchanged \cite{Weiss2015}.

This report exploits exactly this re-parameterisation capability where the same underlying optimisation problem is carried across operations, with only the horizon, constraint set, and cost structure adjusted as the mission moves from one phase to the next.

\subsection{Regulation, Reference Tracking, and Economic Cost}
\label{sec:whyempc-costdesign}

Given that MPC handles constraints natively, the remaining design freedom is
the cost function, and here the RPO-MPC literature reduces to two classical
choices. The first, and by far the most common \cite{Hartley2015,Weiss2015,%
Petersen2023}, penalises the deviation of state and input from a fixed
equilibrium setpoint,
\[
  \ell_{\mathrm{reg}}(x,u) = \|x - x_s\|_Q^2 + \|u - u_s\|_R^2 ,
\]
i.e.\ pure fixed-point regulation. The second replaces the constant
setpoint with a precomputed, time-varying reference $x^\ast_k$ and follows it
in a phase-synchronised manner,
\[
  \ell_{\mathrm{track}}(x_k,u_k) = \|x_k - x^\ast_k\|_Q^2 + \|u_k - u^\ast_k\|_R^2 .
\]
Both retain a quadratic state penalty and differ only in whether the reference
is a point or a trajectory. In this report the closing and docking phases are
point-stabilisation problems, for which $\ell_{\mathrm{reg}}$ is the natural
baseline. The fly-around instead targets a phase-synchronised point on the
NMC. Exactly the trajectory the economic cost is meant to exploit.

An economic cost is especially interesting for RPO because the relative
orbital dynamics admit trajectories that consume no fuel at all yet do not sit
at any fixed point. The CW state-transition matrix has purely imaginary
eigenvalues, and the resulting sinusoidal closed-form solutions yield bounded,
periodic, thrust-free trajectories (NMCs) along which a chaser can remain in
the target's vicinity indefinitely without expending propellant
\cite{Petersen2023}. This exposes the weakness of both quadratic
designs. A fixed-point regulator asked to hold a single point on (or near)
such an orbit is, by construction, fighting the natural drift and burning fuel
to cancel free motion, and while a phase-synchronised tracker can capture
the NMC, it does so only if the fuel-free reference $x^\ast_k$ is known and
supplied in advance.

This is the direct motivation to go with an economic cost.

\subsection{Economic MPC}
\label{sec:whyempc-empc}

Economic MPC (EMPC) lets the cost encode the physical resource of interest
directly \cite{Diehl2011,Faulwasser2018}. For a thrusting spacecraft the
propellant expended over a manoeuvre is the accumulated velocity increment,
\[
  \Delta v = \int \|u(t)\|\,\mathrm{d}t ,
\]
so an exact minimum-propellant objective is the 1-norm stage cost
$\ell \propto \|u\|$. In this work we instead adopt the quadratic
control-effort surrogate
\[
  \ell_{\mathrm{eco}}(u) = \|u\|_R^2 ,
\]
a smooth, strictly convex proxy that keeps each receding-horizon step a
well-conditioned QP and yields continuous thrust profiles, at the price of not
minimising $\Delta v$ exactly. What matters for the present argument is the feature both
share, namely the absence of any state-tracking term. Reported $\Delta v$
values are consequently evaluated a posteriori as $\int\|u\|\,\mathrm{d}t$,
not as the optimised quantity. By omitting the state penalty the optimiser is
free to exploit whichever admissible motion minimises cumulative effort. For a
system whose unforced dynamics already contain the fuel-free NMC trajectories,
this is precisely the freedom a quadratic tracking or regulation cost forfeits.

This freedom, however, forfeits the standard stability certificate. The
economic stage cost carries no state penalty, so it does not vanish only at
the desired steady state and grow away from it, and it therefore cannot serve
as a Lyapunov function. The Riccati-based terminal ingredient that certifies
quadratic MPC through its equivalence to the infinite-horizon LQR
\cite{Hartley2015,Weiss2015} no longer applies. Indeed, an economic stage cost
need not be positive definite with respect to the target at all
\cite{Faulwasser2018}. Recovering a convergence guarantee thus calls for a
different structural property. A well-established route is  the concept of strict
dissipativity \cite{willems1972dissipative}.

A system $x^{+} = f(x,u)$ with stage cost $\ell$ is strictly dissipative with respect to an optimal steady state $(x_s,u_s)$ if there exists a storage function $\lambda:\mathbb{X}\to\mathbb{R}$ and a class-$\mathcal{K}_\infty$ function $\rho$ such that the rotated stage cost
\[
  \tilde\ell(x,u) = \ell(x,u) - \ell(x_s,u_s) + \lambda(x) - \lambda\big(f(x,u)\big)
  \;\ge\; \rho\big(\|x-x_s\|\big) \;>\; 0
  \quad (x \neq x_s),
\]
for all admissible $(x,u)$ \cite{willems1972dissipative,Diehl2011,Angeli2012}. Because $\tilde\ell$ differs from $\ell$ only by a telescoping term and a
constant, both yield the same optimal inputs, yet $\tilde\ell$ is positive definite in $x$ about $x_s$, so the optimal value
function of the rotated problem is a valid Lyapunov function \cite{Angeli2012}. This
restores the standard MPC convergence argument for the economic problem, which
is why constructing the storage function $\lambda$ is the central analytical
step of any EMPC stability claim.

For the CW system under a pure propellant cost $\ell(x,u) = \|u\|$ (or $\|u\|^2$), this necessity is not given due to structural reasons.
The economic cost is minimised, with value zero, by $u = 0$. Since the entire V-bar axis $\{x = [\,0,\,y,\,0,\,0,\,0,\,0\,]\}$ consists of relative equilibria, and because every NMC is a bounded, thrust-free periodic solution of zero stage cost, the optimiser is thus presented with a continuum of zero-cost steady states and periodic orbits. With all of them sharing the minimal average cost, namely $u = 0$.
Strict dissipativity, however, demands a strictly positive margin $\rho(\|x - x_s\|)$ that grows as the state moves away from a single distinguished optimal steady state $x_s$. A continuum of equally optimal, zero-cost solutions yields that no storage function can render the rotated cost uniquely minimal at a single point.

The results of this degeneracy are obvious. A bare-fuel objective contains no mechanism for selecting the desired state or orbit. Thus the controller will drift towards whichever equilibrium is cheapest to reach (see Section~\ref{sec:op2}).

Accordingly, the remainder of this report is about how to repair this degeneracy. The candidate routes all reintroduce, in one form or another, the ``preference'' that the bare economic cost lacks, e.g. hard terminal constraints that pin the target state or orbit directly, terminal costs that shape the value function near the target without forcing it exactly or a stage-cost regularisation that adds a small state-dependent term to recover a strict (or practical) dissipativity inequality. Section~\ref{sec:mpc} formalises the shared machinery behind them, and Sections~\ref{sec:op1}--\ref{sec:op3} examine, operation by operation, which route is appropriate and why.

\section{MPC Formulation \& Concepts}
\label{sec:mpc}

\subsection{System Model}

With state $x = [x,y,z,\dot{x},\dot{y},\dot{z}]^\top$ and control input $u$ (specific thrust, i.e. acceleration), the continuous-time CWH dynamics are \cite{ClohessyWiltshire1960,schaub2018analytical}

\begin{equation}
  \dot{x} = A_c x + B_c u, \qquad
  A_c = \begin{bmatrix}
      0 & 0 & 0 & 1 & 0 & 0 \\
      0 & 0 & 0 & 0 & 1 & 0 \\
      0 & 0 & 0 & 0 & 0 & 1 \\
      3n^2 & 0 & 0 & 0 & 2n & 0\\
      0 & 0 & 0 & -2n & 0 & 0 \\
      0 & 0 & -n^2 & 0 & 0 & 0
  \end{bmatrix} \quad
 B_c = \begin{bmatrix}
     0 & 0 & 0 \\
     0 & 0 & 0 \\
     0 & 0 & 0 \\
     1 & 0 & 0 \\
     0 & 1 & 0 \\
     0 & 0 & 1 \\
 \end{bmatrix}
\end{equation}
where $n$ is the target's mean motion. For the simulation, an ISS-like reference orbit was used, with $n = \SI{0.001133}{rad/s}$ and orbital period $T = \SI{5545}{s}$.

ZOH discretisation at scenario-appropriate sample time $\Delta t$ yields the discrete-time system

\begin{equation}
x_{k+1} = A_d x_k + B_d u_k, \qquad A_d = e^{A_c \Delta t}, \quad B_d = \int_0^{\Delta t} e^{A_c \tau} d\tau \, B_c.
\end{equation}
The terminal-cost matrix $P$ is the solution of the Discrete Algebraic Riccati Equation (DARE)
\begin{equation}
  P = Q + A_d^\top P A_d - A_d^\top P B_d\bigl(R + B_d^\top P B_d\bigr)^{-1} B_d^\top P A_d.
  \label{eq:dare}
\end{equation}

\subsection{Condensed QP Formulation}
\label{sec:condensed-qp}

All controllers in this report are implemented in condensed (single-shooting)
form, in which the predicted states are eliminated and the quadratic program
(QP) is solved over the input sequence alone
\cite{Rawlings2017,Borrelli_Bemporad_Morari_2017}. Substituting the discrete
dynamics $x_{i+1} = A_d x_i + B_d u_i$ repeatedly over a horizon $N$ gives the
lifted prediction
\begin{equation}
  \mathbf{X} = S_x\, x(k) + S_u\, U, \qquad
  \mathbf{X} = [x_1^\top,\dots,x_N^\top]^\top, \quad
  U = [u_0^\top,\dots,u_{N-1}^\top]^\top,
\end{equation}
with prediction matrices
\begin{equation}
  S_x = \begin{bmatrix} A_d \\ A_d^2 \\ \vdots \\ A_d^N \end{bmatrix},
  \qquad
  S_u = \begin{bmatrix}
    B_d & 0 & \cdots & 0 \\
    A_d B_d & B_d & \cdots & 0 \\
    \vdots & & \ddots & \vdots \\
    A_d^{N-1}B_d & \cdots & A_d B_d & B_d
  \end{bmatrix}.
  \label{eq:pred-matrices}
\end{equation}
For a quadratic stage and terminal cost with block-diagonal weights
$\bar{Q} = \operatorname{blkdiag}(Q,\dots,Q,P)$ and $\bar{R} = I_N \otimes R$,
and an optional stacked reference $\mathbf{R}$, the finite-horizon problem
reduces to the strictly convex QP
\begin{equation}
  \min_{U}\ \tfrac12 U^\top H U + f^\top U
  \quad\text{s.t.}\quad
  \underbrace{-u_{\max}\mathbf{1} \le U \le u_{\max}\mathbf{1}}_{\text{input}},\quad
  \underbrace{A_{\mathrm{in}} U \le b_{\mathrm{in}}}_{\text{state / geometric}},\quad
  \underbrace{A_{\mathrm{eq}} U = b_{\mathrm{eq}}}_{\text{terminal}},
  \label{eq:qp-full}
\end{equation}
with
\begin{equation}
  H = 2\bigl(S_u^\top \bar{Q}\, S_u + \bar{R}\bigr),
  \qquad
  f = 2\, S_u^\top \bar{Q}\,(S_x\, x(k) - \mathbf{R}),
  \label{eq:qp-reg}
\end{equation}
where $\mathbf{R} = 0$ for pure regulation.

Any state constraint the chaser must respect at every time step, written
generally as $C\,x_i \le d$, can be rewritten purely in terms of the inputs
$U$ instead of the states, using the same substitution
\eqref{eq:pred-matrices} that built the prediction in the first place. Collecting
all such constraints together gives the compact form
$A_{\mathrm{in}}U \le b_{\mathrm{in}}$ used in the QP.

Two such constraints are used to shape the approach geometry. (i) An
along-track half-space, $y_i \ge 0$, simply keeps the chaser on the
correct side of the docking port throughout the approach. (ii) A
line-of-sight cone of half-angle $\alpha$ around the $+$V-bar
axis, $\sqrt{x_i^2+z_i^2}\le y_i\tan\alpha$, keeps the chaser inside a cone
so the docking sensors always have a clear view of the target. This cone is
a smooth, curved shape, which the QP solver cannot handle directly, so it is
approximated by $M$ flat faces,
\begin{equation}
  \cos\theta_j\,x_i + \sin\theta_j\,z_i
  \;\le\; \tan\alpha\,\cos(\tfrac{\pi}{M})\,y_i,
  \qquad \theta_j = \tfrac{2\pi j}{M},\ j=0,\dots,M-1,
  \label{eq:los-cone}
\end{equation}
where the extra $\cos(\pi/M)$ factor shrinks the faces slightly inward so
that the resulting polygon sits entirely inside the true cone, never
outside it \cite{Weiss2015}. This way the approximation is conservative. Any trajectory satisfying the flat-face version also satisfies the real
cone constraint.

The condition placed on the very last predicted state is built the same
way. Pinning the chaser exactly to the docking state at the end of the
horizon, $x_N = x_{\mathrm{dock}}$, becomes an equation in $U$ alone,
while a softer version, only requiring the chaser to land somewhere inside
a small box around the target at the end of the horizon, is written as
$\|C_p x_N\|_\infty \le r$. A perfectly round target region would again
turn the problem into a curved, more expensive one to solve, so a box is
used instead, chosen so it always sits inside the true target region.

The QP is solved with MATLAB's \texttt{quadprog} \cite{MATLABquadprog}. Only
the first block $u^\star(k) = U^\star_{1:n_u}$ is applied and the horizon
recedes one step. Real-time and embedded implementations of this dense QP are
well established \cite{WANG20086974,Jerez2011}. This construction is shared by
every controller below, which differ only in the stage cost, terminal
ingredient, reference, and active constraint set.

\subsection{Economic MPC Formulation}
\label{sec:empc-reg}

Every economic controller sets $Q = 0$ in the condensed QP of Section~\ref{sec:condensed-qp}, so the stage cost $\ell(x,u) = \|u\|^2$ reflects propellant directly. The structural consequences of this choice, the dissipativity gap and the design axis for closing it, are developed in the context of Operation 1 where they first arise (Section~\ref{sec:op1}).
The economic regulation controller replaces the quadratic state-tracking stage cost with a fuel-only cost while retaining the DARE terminal cost:
\begin{subequations}
\label{eq:empc-reg}
\begin{align}
  \min_{U}\ & \sum_{i=0}^{N-1} u_i^\top R_{\mathrm{eco}}\, u_i + x_N^\top P\, x_N \\
  \text{s.t.}\ & x_{i+1} = A_d x_i + B_d u_i, \quad -u_{\max} \le u_i \le u_{\max}, \quad x_0 = x(k).
\end{align}
\end{subequations}
This is the regulation MPC \eqref{eq:qp-reg} with $Q_{\mathrm{stage}} = 0$ and $R = R_{\mathrm{eco}} = I_{n_u}$. The terminal cost $P$ and gain $K$ are computed from the DARE \eqref{eq:dare} with stabilising weights $(Q_{\mathrm{dare}} \succ 0,\, R_{\mathrm{dare}})$, $A_{\mathrm{cl}} := A_d - B_d K$ Schur. Following \cite{AMRIT2011178}, $V_f(x) = x^\top Px$ is a valid terminal cost for the economic stage cost provided the descent condition $P - A_{\mathrm{cl}}^\top P A_{\mathrm{cl}} \succeq K^\top R_{\mathrm{eco}} K$ holds; using the DARE identity $P - A_{\mathrm{cl}}^\top P A_{\mathrm{cl}} = Q_{\mathrm{dare}} + K^\top R_{\mathrm{dare}} K$, this reduces to
\begin{equation}
\label{eq:m1}
  M_1 := Q_{\mathrm{dare}} + K^\top(R_{\mathrm{dare}} - R_{\mathrm{eco}})K \;\succeq\; 0,
\end{equation}
Condition~\eqref{eq:m1} holds with margin at every sampling interval used in
this report. Table~\ref{tab:m1} lists the eigenvalues of $M_1$ for the closing
and docking scenarios.
\begin{table}[h]
\centering
\caption{Eigenvalues of $M_1$~\eqref{eq:m1} at each scenario's sampling
interval}
\label{tab:m1}
\begin{tabular}{lcc}
\toprule
Scenario & $\Delta t$ & $\operatorname{eig}(M_1)$ \\
\midrule
Closing & 120\,s & $\{0.0100,\,0.0100,\,0.0100,\,3.125,\,3.127,\,3.152\}$ \\
Docking & 10\,s  & $\{0.0104,\,0.0104,\,0.0104,\,17.53,\,17.53,\,17.59\}$ \\
\bottomrule
\end{tabular}
\end{table}
The QP matrices are
\begin{equation}
  H = 2\bigl(S_u^\top \bar{P}\, S_u + \bar{R}_{\mathrm{eco}}\bigr),
  \qquad
  f = 2\, S_u^\top \bar{P}\, S_x\, x(k),
\end{equation}
with $\bar{P} = \operatorname{blkdiag}(\underbrace{0,\dots,0}_{N-1}, P)$ and $\bar{R}_{\mathrm{eco}} = I_N \otimes R_{\mathrm{eco}}$.\\
At each sampling instant the QP is solved in error coordinates
$e_k = x(k) - x_{\mathrm{ref}}$ (the docking port $x_{\mathrm{dock}}$ in
Operation~3, the handover point $x_{\mathrm{switch}}$ in Operation~1), the first
input $u^\star(k) = U^\star_{1:n_u}$ is applied, the state is propagated one
step, and the horizon recedes \cite{MAYNE2000789,Rawlings2017}.
During fly-around the chaser acquires and maintains a Natural Motion
Circumnavigation (NMC). Its full
state
\begin{equation}
  x^*_{\mathrm{nmc}}(t) =
  \begin{bmatrix}
    b\cos(\phi_0 + n t) \\
    y_c - 2b\sin(\phi_0 + n t) \\
    0 \\
    -b\, n \sin(\phi_0 + n t) \\
    -2b\, n \cos(\phi_0 + n t) \\
    0
  \end{bmatrix},
  \label{eq:nmc-ref}
\end{equation}
has radial semi-axis $b$, initial phase $\phi_0$, along-track drift centre $y_c$, and mean motion $n$.
Unlike the regulation controllers, this operation uses no terminal cost,
because a soft terminal penalty alone leaves the target orbit unselected under a
pure-fuel cost. Instead the condensed QP keeps the fuel-only stage cost
($H = 2 I_{n_u N}$, $f = 0$) and pins the orbit in one of two ways. Sampling the
reference at the controller rate, $x^*_j = x^*_{\mathrm{nmc}}(j\,\Delta t)$ for
$j = 0,\dots,P-1$ with $P = T/\Delta t$ samples per period, the first variant
imposes a phase-synchronised terminal equality on all six states,
\begin{equation}
  x_N = x^*_{\sigma(k)}, \qquad
  \sigma(k) = \bigl((k + N - 1)\bmod P\bigr) + 1,
  \label{eq:nmc-termeq}
\end{equation}
while the second replaces it with an $\varepsilon$-regularised phase-synchronised
stage cost $\ell_{\mathrm{aug}} = \|u\|^2 + \varepsilon\|x - x^*_k\|^2$ that
recovers strict $P$-periodic dissipativity without any terminal constraint
\cite{Zanon2017,MULLER2016128,Koehler2018}. In both cases the phase index
$\sigma(k)$ is a deterministic function of the global controller step $k$. The
reference is phase-locked to mission time, advancing one sample per step and
wrapping every $P$ steps. The details are
developed in Section~\ref{sec:op2}.

\subsection{Software Availability}
\label{subsec:sw_availability}

The implementation developed for this work is available as an open-source repository under the BSD 3-Clause License:\\
\url{https://github.com/maiergamer9/empc-rpo-thesis}\\
The repository contains the controller implementation, simulation scripts, and supporting material required to reproduce the numerical studies presented in Section~\ref{sec:operations}.

\section{Mission Operations}
\label{sec:operations}

\subsection{Operation 1: Far-Range Closing}
\label{sec:op1}

Operation 1 is the closing transfer from a \SI{10}{km} V-bar hold to a handover
box at \SI{300}{m}, run at $\Delta t = \SI{120}{s}$, $N = 20$ and with thrust bound
$u_{\max} = \SI{e-2}{m/s^2}$.
The ADRIOS ConOps initiates rendezvous near \SI{30}{km} and flies the far-range
approach with ground-planned, impulsive burns under absolute navigation \cite{Woicke2025}. That
segment is governed by orbital-mechanics targeting rather than closed-loop
proximity control and lies outside the scope of this study. The present work
models the close-proximity segment, which begins around \SI{10}{km} once
relative navigation is available and autonomous, continuously actuated MPC takes
over. The actuator is accordingly modelled
as a low-thrust device delivering a bounded acceleration
$\|u\|_\infty \le u_{\max}$, held piecewise-constant over each sample by the
zero-order hold, as opposed to the impulsive-$\Delta v$ model of the far-range phase.
The simulation stops as soon as the chaser enters a \SI{50}{m} ball around the
switch-point, i.e.\ when the full-state error satisfies
$\|x_k - x_{\mathrm{switch}}\|_2 < \SI{50}{m}$. This suffices because Operation 2 does
not require its initial condition to be exactly
$x_0 = [0,\,300,\,0,\,0,\,0,\,0]^\top$. The horizon is capped at five orbital
periods ($t_{\max} = 5T$) in which every closing run converges well within,
so five periods is the maximum admissible duration, not the realised one.

\subsubsection{Dissipativity Gap and the Design Axis}

The initial state $x_0 = [0,\,10\,000,\,0,\,0,\,0,\,0]^\top$ is itself a
zero-input CW equilibrium. Under the pure fuel cost $\ell(x,u) = \|u\|^2$, the
optimal first move is to do nothing and the closed loop never leaves
\SI{10}{km}. The value function remains bounded below by zero, but it is not
positive definite with respect to the target. It attains its minimum, zero, at
every V-bar equilibrium, including the \SI{10}{km} initial point,
rather than uniquely at the desired handover state. Equivalently, the system is not strictly dissipative
with respect to the target regime \cite{GRUNE2013725}. No storage function
renders the rotated stage cost positive definite, and the receding-horizon
optimum is simply $u \equiv 0$. This is the core requirements finding for this
operation. A pure fuel stage cost is insufficient for far-range closing
rendezvous, and at least one additional structural ingredient (a terminal
constraint, a terminal cost, or a state regularisation) is needed.

Closing this gap requires injecting the location and tail value of the desired steady state through a terminal ingredient. Three channels for doing so define the design axis for Operation 1:

\begin{itemize}
\item \textbf{Run~A1}: A terminal ball constraint $\|C_p e_N\|_\infty \le r$, with $r = \SI{200}{m}$, that injects the target location (the switch-point) as a hard feasibility requirement, leaving the fuel stage cost untouched. Finite-time arrival of the predicted position into the handover box is observed in simulations.
\item \textbf{Run~B}: A stage-cost regularisation $\ell = \|u\|^2 + \rho_c \|e\|^2$ , with $\rho_c = 10^{-11}$, that tilts the cost landscape so the target becomes the unique minimiser, with no offline computation. Yields practical asymptotic stability, backed by \cite{AMRIT2011178}'s stability result. The convergence neighbourhood shrinks as $N \to \infty$ or $\rho_c \to \infty$, but a larger $\rho_c$ distorts the fuel objective.
\item \textbf{Run~C}: A CLF terminal cost $V_f(e) = e^\top P e$ with
  $M_1 \succeq 0$ (\S\ref{sec:empc-reg}), verified numerically at
  $\Delta t = \SI{120}{s}$ (Table~\ref{tab:m1}). By \cite{AMRIT2011178}, the
  descent property $M_1 \succeq 0$ certifies the asymptotic
  average-performance bound (the closed-loop running cost is asymptotically
  no worse than the optimal steady state) but it does not certify
  asymptotic stability of the target. Asymptotic stability would additionally
  require strict dissipativity, which again fails here for the same reasons already discussed in Section~\ref{sec:op1}. Convergence to the target is observed in
  simulation and is therefore reported as an empirical result, not a
  proven guarantee.
\end{itemize}

\subsubsection{Results}

\begin{figure}[t]
  \centering
  \resizebox{1\textwidth}{!}{\input{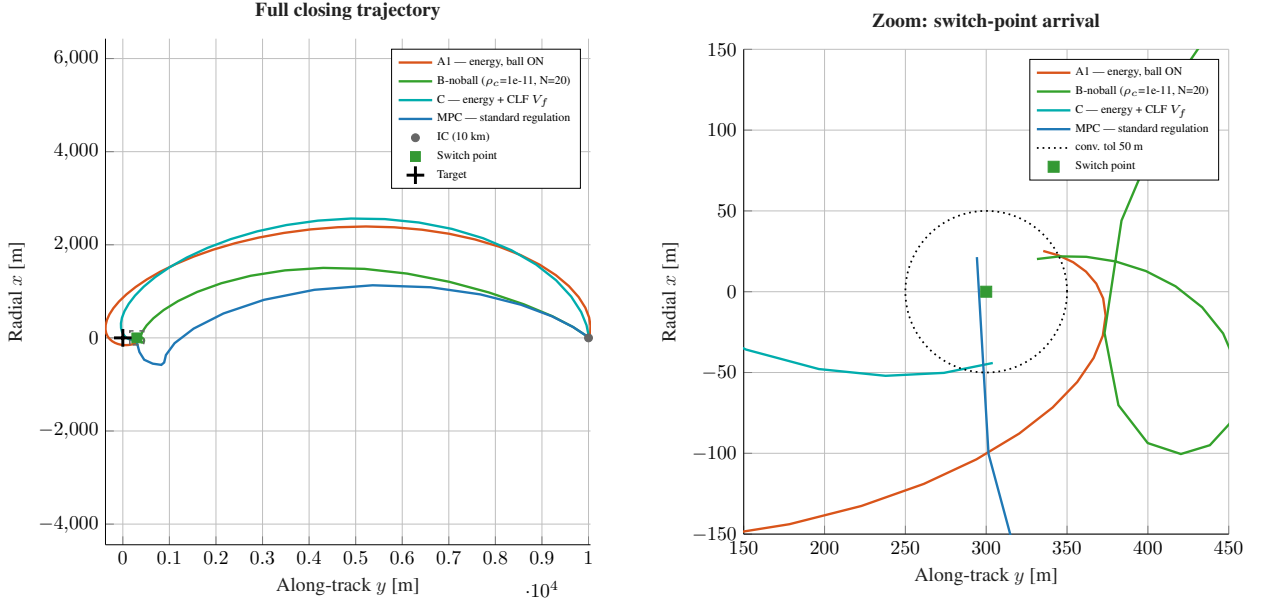}}
  \caption{Closing trajectories --- full view and switch-point zoom.}
  \label{fig:s4-hill}
\end{figure}

The three EMPC channels are compared against a Standard (quadratic-tracking) MPC
with weights $Q_{\mathrm{dare}} = \mathrm{diag}(10^{-2},10^{-2},10^{-2},1,1,1)$
and $R_{\mathrm{dare}} = \mathrm{diag}(10^{4},10^{4},10^{4})$. Table~\ref{tab:s4}
and Fig.~\ref{fig:s4-hill} report the outcome.
All three channels resolve the $u\equiv 0$ degeneracy and complete the closing
at a fraction of the Standard-MPC fuel cost. Standard MPC spends
\SI{31.72}{m/s}, whereas the economic variants range from \SI{3.29}{m/s}
(Run~A1) to \SI{18.33}{m/s} (Run~B). The ranking $\text{A1} < \text{C} < \text{B}$ mirrors how strongly each
ingredient distorts the pure-fuel objective. The terminal ball (A1) leaves the
stage cost untouched and is cheapest while the CLF terminal cost (C) adds only a
terminal penalty and the stage-cost regularisation (B), which penalises state
error at every step, is the most expensive economic option while still
undercutting Standard MPC by \SI{42}{\percent}. That the hard terminal ball is
the least costly way to inject the target confirms the requirements reading of
\S\ref{sec:op1}.
The saving is bought with time. Driven by its state-tracking cost, Standard MPC
converges fastest (21 steps, \SI{0.67}{h}) by thrusting hard, whereas the
fuel-optimal Run~A1 coasts on the natural drift and takes three times as long
(61 steps, \SI{2.0}{h}). Runs~B and~C fall in between ($\approx\!\SI{1.2}{h}$).
This fuel-versus-time trade is the central engineering choice for the closing
phase.

\begin{table}[h]
\centering
\caption{S4 closing-approach run comparison.}
\label{tab:s4}
\begin{tabular}{lccccc}
\toprule
Run & $\Delta v$ [m/s] & Steps / time to conv. & term. distance [m]\\
\midrule
Standard MPC & 31.7196 & 21 / 0.67 h & 22.17 \\
Run A1 (EMPC + terminal ball) & 3.2888 & 61 / 2.00 h & 43.36  \\
Run B ($\rho_c$ regularisation) & 18.3321 & 36 / 1.17 h & 37.32 \\
Run C (EMPC + CLF) & 5.9800 & 37 / 1.20 h & 44.29  \\
\bottomrule
\end{tabular}
\end{table}

\subsection{Operation 2: NMC Fly-Around}
\label{sec:op2}

Operation 2 acquires a $2{:}1$ NMC fly-around orbit of semi-axis \SI{75}{m} from the \SI{300}{m} handover point, run at $\Delta t = T/92 \approx \SI{60.27}{s}$, $N = 92$ (a full orbital period). The economic degeneracy here is sharper than in Operation 1. Because every passive NMC orbit is a zero-cost trajectory, pure fuel cost leaves the controller indifferent among the entire one-parameter family of ellipses, and it simply coasts on its natural drift ($\Delta v = 0$, no orbit acquired). Critically, the periodic analogue of strict dissipativity also fails for the pure-fuel cost on this orbit family, so the rotated-cost route available in principle for economic MPC is closed here \cite{Zanon2017,MULLER2016128}.
A sequence of failure-mode runs (Table~\ref{tab:op2_modes}) establishes that neither a state cost nor a range band substitutes for the missing terminal ingredient. The instructive case is Run~3a. $Q$ combined with a band manufactures a continuum of equal-cost non-equilibrium states and locks the QP into an expensive powered limit cycle, strictly worse than pure fuel alone. The ingredient that both selects the target orbit $\Pi^*$ and fixes its phase is a hard periodic terminal equality $x_N = x^*_\varphi$ tying the horizon end to the phase-consistent point on the orbit.

\begin{table}[h]
\centering
\caption{Operation 2 --- outcome by cost / constraint combination.}
\label{tab:op2_modes}
\begin{tabular}{lllll}
\toprule
Run & Cost & Constraint & Outcome \\
\midrule
1  & fuel & none & coasts, no orbit  \\
2  & fuel & periodic terminal eq. & captures $\Pi^*$  \\
3c & $Q$  & none & regulates to origin  \\
3b & fuel & band & V-bar equilibrium  \\
3a & $Q$  & band & powered limit cycle  \\
4  & fuel & IC on $\Pi^*$ & captures $\Pi^*$ \\
5  & phase-synchronised $\ell_{\mathrm{aug}}$ ($\varepsilon = 10^{-4}$)  & none & captures $\Pi^*$ \\
\bottomrule
\end{tabular}
\end{table}
The periodic terminal equality captures $\Pi^*$ at total $\Delta v = 0.071\,\text{m/s}$,
with the target geometry. It discovered semi-axis $b \approx 74.9\,\text{m}$ against the
$75\,\text{m}$ target (range ratio 2.004), terminal control decaying to
$\|u\| \to 3.6 \times 10^{-9}\,\text{m/s}^2$ and last-orbit phase error at the
sub-decimetre level. A regularised alternative (Run 5) drops the terminal constraint
entirely, restoring strict periodic dissipativity through the phase-synchronised
augmented cost $\ell_{\text{aug}}$ ($\varepsilon = 10^{-4}$), reaching the same orbit
($b = 75.0\,\text{m}$, ratio 2.000) at total $\Delta v = 7.519\,\text{m/s}$. The two routes converge to the same $\Pi^*$. The terminal equality is exact but binds the horizon to a full period and adds six equality rows. The regularised route frees the horizon and drops the constraint at the cost of a sub-\si{cm/s} fuel bias.
Both implementations are compared to a standard Tracking MPC with $Q = I_6$, $R = I_3$ and terminal cost
$P = \operatorname{DARE}(A_d,B_d,Q,R)$, over a horizon $N_{\mathrm{track}}=20$.

\begin{table}[h]
  \centering
  \caption{NMC orbit acquisition from the shared V-bar hold
    $x_0=[0,300,0,0,0,0]^\top$: standard tracking MPC versus the economic
    controllers. Capture = first time the phase error stays below $5$\,m.}
  \label{tab:s2-tracking-comparison}
  \begin{tabular}{lrrr}
    \toprule
    Controller & $\Delta v$ [m/s] & Saving vs.\ tracking & Capture [orbits] \\
    \midrule
    Standard tracking MPC              & 12.217 & ---     & 0.08 \\
    Run~2 (terminal eq.)               &  0.071 & 99.4\%  & 2.14 \\
    Run~5 (periodic $\ell_{\mathrm{aug}}$) &  7.519 & 38.5\%  & 0.08 \\
    \bottomrule
  \end{tabular}
\end{table}

\begin{figure*}[t]
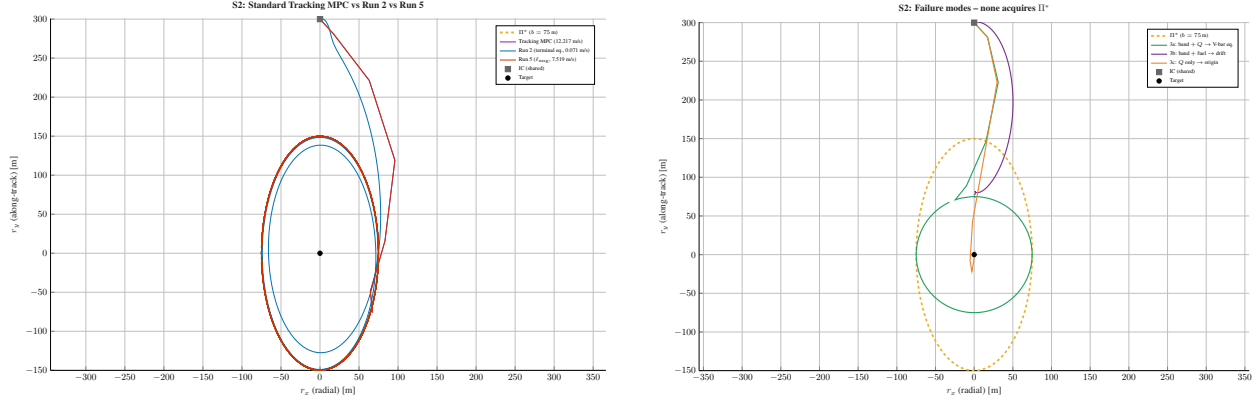

  \centering
  \begin{subfigure}[t]{0.48\linewidth}
    \centering
    \resizebox{\linewidth}{!}{\input{figures/tracking_vs_run2_run5.tikz}}
    \caption{Standard tracking MPC vs.\ the economic Run~2 (periodic terminal
      equality) and Run~5 (periodic $\ell_{\mathrm{aug}}$).}
    \label{fig:s2-tracking-vs-run2-run5}
  \end{subfigure}
  \hfill
  \begin{subfigure}[t]{0.48\linewidth}
    \centering
    \resizebox{\linewidth}{!}{\input{figures/failure_modes_overlay.tikz}}
    \caption{Failure modes: neither a state band nor a state cost alone
      selects the NMC orbit.}
    \label{fig:s2-failure-modes}
  \end{subfigure}
  \caption{Fly-around orbit acquisition and failure-mode comparison.}
  \label{fig:s2-combined}
\end{figure*}

\subsection{Operation 3: Final Docking}
\label{sec:op3}

Operation 3 is the terminal docking approach from \SI{150}{m} to the capture
envelope, run at $\Delta t = \SI{10}{s}$, $N = 24$,
$u_{\max} = \SI{e-2}{m/s^2}$, from a V-bar forced-keep-point initial
condition, with convergence declared at $\|e_r\| < \SI{1}{m}$ on the
position error. An appropriate distance for switching from GNC to the
capture mechanism. A pure-fuel stage cost would again yield the degeneracy
described in \ref{sec:op1} and \ref{sec:op2}. Thus, Operation 3 again
implements an EMPC with terminal cost term and a Standard MPC, with the same
weight matrices as in Operation 1. Instead of a regularised stage-cost or a
terminal ball, the objective of this phase admits a hard terminal equality
constraint $x_{N} = x_{\mathrm{dock}}$. All three are compared in $\Delta v$,
convergence time and terminal distance.

Neither the \SI{150}{m} start nor the horizon $N=24$ is a free parameter and
both are imposed by the upstream mission. The fly-around of Operation~2
departs at the V-bar crossing of the natural motion circumnavigation, which
for a radial semi-axis $b = \SI{75}{m}$ lies at $y = 2b = \SI{150}{m}$. This
geometry in turn sets a hard requirement on the docking controller. The
terminal-equality constraint $x_N = x_{\mathrm{dock}}$ demands that the
origin be reachable in both position and velocity within the horizon.
The minimum feasible horizon was established empirically. $N$ was swept
downward in simulation until the docking QP first became infeasible in the
first steps, at which point the controller silently degrades to a
best-effort, constraint-violating fallback. $N=24$ is the minimum feasible
horizon found this way, and is used as the common horizon for all three
runs. A closed-form derivation of this bound for the coupled, constrained
CWH dynamics is
left as future work. The requirement binds only Run~3. Runs~1 and~2
replace the equality with a terminal cost and stay feasible at any $N$, but
since a common horizon is used, $N=24$ becomes the design point, fixed by the
most demanding controller. This is a concrete instance of a
mission-derived MPC requirement.
 
\begin{figure}[htbp]
  \centering
  \resizebox{\linewidth}{!}{\input{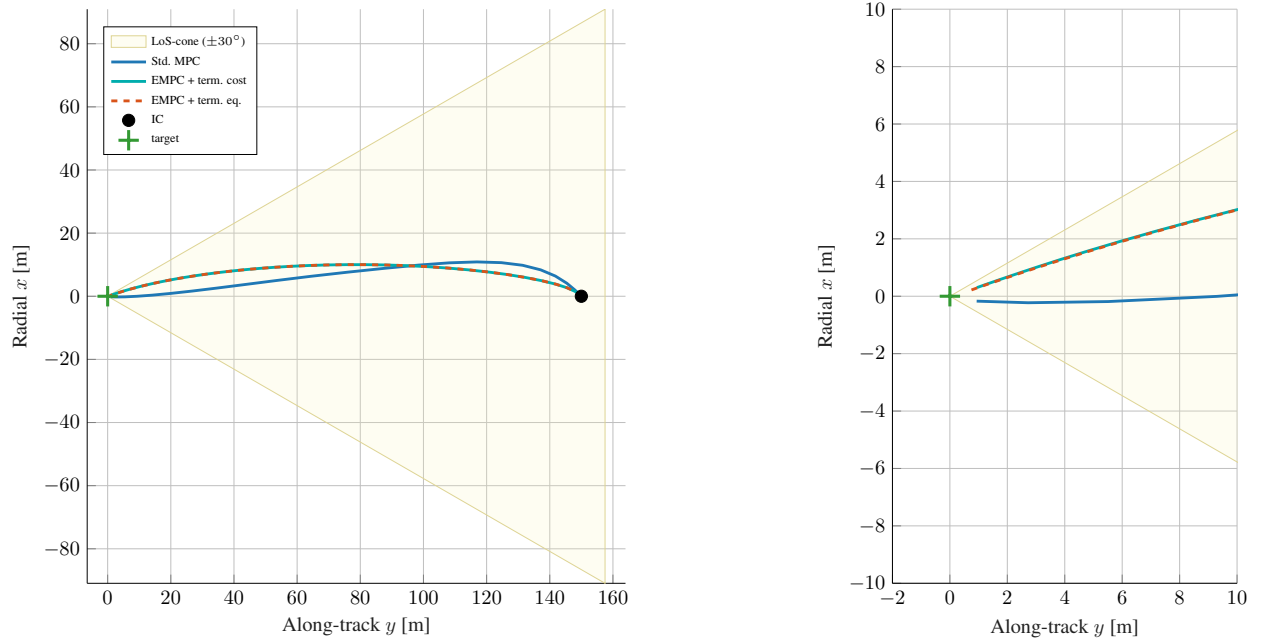}}
  \caption{Hill-frame docking trajectory (radial $x$ vs.\ along-track $y$).
    Left: full approach from the \SI{150}{m} V-bar hold to the port; right:
    terminal zoom ($\leq \SI{10}{m}$).}
  \label{fig:s1_hill}
\end{figure}

The approach is deliberately confined to an along-track corridor. In
particular a line-of-sight cone of half-angle $30^\circ$ about the $+$V-bar
axis, together with a $y \geq 0$ half-space, constrains the chaser to close
on the port from the along-track direction. This is the geometry preferred
for docking-port alignment and capture-sensor line-of-sight, and it keeps the
chaser out of the radial band from which a dispersion would carry it across
the port. Constrained MPC for the terminal docking approach has been studied
extensively \cite{Hartley2015,Weiss2015,DiCairano2012}. Table~\ref{tab:s1}
reports the three-run comparison and Fig.~\ref{fig:s1_hill} shows the
Hill-frame trajectories with the corridor shaded.
 
\begin{table}[h]
\centering
\caption{S1 docking-approach run comparison (\SI{150}{m}, $N=24$).}
\label{tab:s1}
\begin{tabular}{lccc}
\toprule
Run & $\Delta v$ [m/s] & Steps / time to conv. & Term.\ distance [m] \\
\midrule
Run 1 -- Standard MPC                   & 2.451 & 24 / 4.00 min & 0.948 \\
Run 2 -- EMPC + CLF                     & 1.314 & 43 / 7.17 min & 0.988 \\
Run 3 -- EMPC + hard terminal equality  & 1.322 & 44 / 7.33 min & 0.691 \\
\bottomrule
\end{tabular}
\end{table}
 
EMPC roughly halves the fuel cost relative to Standard MPC
(\SI{1.31}{m/s} vs.\ \SI{2.45}{m/s}, a $46.4\%$ saving) at the cost of a longer settling
time (43 vs.\ 24 steps, \SI{7.2}{min} vs.\ \SI{4.0}{min}). Runs~2 and~3 are
near-identical in $\Delta v$ (\SI{1.31}{m/s} vs.\ \SI{1.32}{m/s}). Since the
V-bar hold is a CW equilibrium, the CLF terminal cost $P_{\mathrm{clf}}$
already acts as an implicit soft terminal equality, so the explicit
constraint of Run~3 adds little fuel. Its effect is instead on the terminal
distance, which it tightens from \SI{0.99}{m} to \SI{0.69}{m}.
 
Declaring convergence at the \SI{1}{m} capture scale is also what keeps the
corridor constraints well-posed. The trajectories halt below \SI{1}{m},
upstream of the cone apex and the $y = 0$ face, where the cone and half-space
would otherwise become singular.

\newpage

\section{Cross Table}
\label{sec:cross}

The three operations were analysed separately, each with its own dissipativity
gap, terminal ingredient, and certificate. Table~\ref{tab:cross-sweep} first
combines them end-to-end, giving the total fuel and time cost of running the
mission with every controller choice per operation, so the cumulative effect
of stacking ingredients across the sequence is visible in one place. Table~\ref{tab:cross}
then draws the requirements analysis itself together. For each operation it
lines up the mission requirement that motivated the ingredient, what that
ingredient formally guarantees, what was actually observed in simulation, and
the residual limitation that guarantee does not cover.

\begin{table}[h]
\centering
\caption{End-to-end mission cost for every controller combination
  (Closing $\times$ Fly-around $\times$ Docking). Total $\Delta v$ excludes the
  fixed departure burn; duration is for the complete three-operation
  sequence, computed from the respective initial conditions of each
  operation.}
\label{tab:cross-sweep}
\begin{tabular}{clllrr}
\toprule
\# & Closing & Flyaround & Docking & $\Delta v_{\mathrm{tot}}$ [m/s] & Duration [h] \\
\midrule
1  & energy+ball (A1)     & periodic-eq (R2)  & EMPC-CLF (R2)    & 4.7692  & 11.01 \\
2  & energy+ball (A1)     & periodic-eq (R2)  & EMPC-termeq (R3) & 4.7771  & 11.01 \\
3  & energy+ball (A1)     & augmented (R5)    & EMPC-CLF (R2)    & 11.0724 & 11.01 \\
4  & energy+ball (A1)     & augmented (R5)    & EMPC-termeq (R3) & 11.0809 & 11.01 \\
5  & dissipative (B)      & periodic-eq (R2)  & EMPC-CLF (R2)    & 19.9102 & 10.14 \\
6  & dissipative (B)      & periodic-eq (R2)  & EMPC-termeq (R3) & 19.9103 & 10.14 \\
7  & dissipative (B)      & augmented (R5)    & EMPC-CLF (R2)    & 26.5352 & 10.14 \\
8  & dissipative (B)      & augmented (R5)    & EMPC-termeq (R3) & 26.5431 & 10.15 \\
9  & energy+CLF (C)       & periodic-eq (R2)  & EMPC-CLF (R2)    & 7.6459  & 10.21 \\
10 & energy+CLF (C)       & periodic-eq (R2)  & EMPC-termeq (R3) & 7.6539  & 10.21 \\
11 & energy+CLF (C)       & augmented (R5)    & EMPC-CLF (R2)    & 13.4763 & 10.21 \\
12 & energy+CLF (C)       & augmented (R5)    & EMPC-termeq (R3) & 13.4842 & 10.21 \\
\bottomrule
\end{tabular}

\end{table}

\begingroup
\footnotesize
\renewcommand{\arraystretch}{1.25}
\begin{longtable}{@{}p{0.7cm}p{2.5cm}p{2.7cm}p{3.0cm}p{2.6cm}p{3.0cm}@{}}
\caption{Cross-scenario requirements synthesis: for each operation, the mission
  requirement that drives the MPC ingredient, the theoretical property claimed
  for it in the cited EMPC literature, the observed result, and the residual
  limitation. Baseline (quadratic) controllers in grey for contrast. The
  entries in the ``Claimed property'' column are not re-derived in this
  report. They state the guarantee associated with the cited result under its
  stated hypotheses, a system-specific verification of which is left as
  future work (Section~\ref{sec:conclusion}).}

\label{tab:cross}\\
\toprule
Op & Ingredient (Run) & Mission req. & Claimed property & Observed result & Limitation \\
\midrule
\endfirsthead
\multicolumn{6}{c}{\tablename~\thetable\ (continued)}\\
\toprule
Op & Ingredient (Run) & Mission requirement & Claimed property & Observed result & Limitation \\
\midrule
\endhead
\bottomrule
\endfoot
 
% -------- Operation 1 --------
\multirow{4}{*}{1}
 & \textcolor{lightgray}{Quadratic tracking (Std MPC)}
   & \textcolor{lightgray}{Reach box around \SI{300}{m} switch point }
   & \textcolor{lightgray}{Asymptotic stability (LQR)}
   & \textcolor{lightgray}{$\Delta v=\SI{31.7}{m/s}$, \SI{0.67}{h}}
   & \textcolor{lightgray}{Fights drift; $5$--$10\times$ EMPC fuel} \\
\cmidrule(l){2-6}
 & Pure fuel + $\ell_\infty$ terminal ball, $r=\SI{200}{m}$ (A1)
   &
   & Predicted arrival into box (feasibility); no stability cert.
   & $\Delta v=\SI{3.29}{m/s}$, \SI{2.0}{h} (best fuel)
   & Slowest; feasibility $\ne$ stability \\
\cmidrule(l){2-6}
 & Pure fuel + $\rho_c\|e\|^2$ regularisation (B)
   &
   & Practical asymptotic stability
   & $\Delta v=\SI{18.3}{m/s}$, \SI{1.17}{h}
   & $\rho_c$ distorts fuel; underlying stabilisability hypothesis not verified here \\
\cmidrule(l){2-6}
 & Pure fuel + CLF terminal cost, $M_1\succeq0$ (C)
   &
   & Avg-performance bound (\emph{not} asymptotic stability)
   & $\Delta v=\SI{5.98}{m/s}$, \SI{1.2}{h}
   & Strict dissipativity fails; convergence empirical \\
\midrule
 
% -------- Operation 2 --------
\multirow{3}{*}{2}
 & \textcolor{lightgray}{Quadratic tracking (Std MPC)}
   & \textcolor{lightgray}{Acquire NMC orbit}
   & \textcolor{lightgray}{Asymptotic stability}
   & \textcolor{lightgray}{$\Delta v=\SI{12.2}{m/s}$, cap.\ 0.08 orb.}
   & \textcolor{lightgray}{Fights free motion; high fuel} \\
\cmidrule(l){2-6}
 & Pure fuel + periodic terminal eq.\ $x_N=x^*_{\sigma(k)}$ (Run 2)
   &
   & Recursive feasibility + avg-performance bound; no orbit-stability cert.
   & $\Delta v=\SI{0.071}{m/s}$ ($99.4\%$), $b\approx\SI{74.9}{m}$
   & Horizon bound to full period $N=P=92$; slow capture (2.1 orb.); hypotheses not re-verified for this system \\
\cmidrule(l){2-6}
 & Pure fuel + $\varepsilon\|x-x^*_k\|^2$ phase-sync (Run 5)
   &
   & $P$-periodic strict diss.\ claimed by cited result (constancy of $\ell,V_N$ on $\Pi^*$ assumed, not checked here) $\Rightarrow$ practical stability
   & $\Delta v=\SI{7.52}{m/s}$ ($38.5\%$), $b=\SI{75.0}{m}$
   & Sub-\si{cm/s} fuel bias; not exact \\
\midrule
 
% -------- Operation 3 --------
\multirow{3}{*}{3}
 & \textcolor{lightgray}{Quadratic tracking + DARE (Std MPC)}
   & \textcolor{lightgray}{Reach docking-port vicinity ($<$\SI{1}{m})}
   & \textcolor{lightgray}{Asymptotic stability}
   & \textcolor{lightgray}{$\Delta v=\SI{2.45}{m/s}$, 24 steps}
   & \textcolor{lightgray}{Highest docking fuel; saturated} \\
\cmidrule(l){2-6}
 & Pure fuel + CLF terminal cost (Run 2)
   &
   & Avg-performance bound
   & $\Delta v=\SI{1.31}{m/s}$ ($46\%$), term.\ \SI{0.99}{m}
   & No asymptotic-stability certificate \\
\cmidrule(l){2-6}
 & Pure fuel + hard terminal eq.\ $x_N=x_{\mathrm{dock}}$ (Run 3)
   &
   & Exact terminal state (feasibility); $N=24$ found empirically
   & $\Delta v=\SI{1.32}{m/s}$, term.\ \SI{0.69}{m} (tightest)
   & $N=24$ from sim.\ sweep; analytical derivation is future work \\
\end{longtable}
\endgroup

\section{Conclusion}
\label{sec:conclusion}
 
This report developed and applied a mission-centric EMPC design framework to
three sequential proximity operations of the ADRIOS / ClearSpace-1 mission.
For each operation, the same condensed-QP formulation was instantiated with a
pure-fuel stage cost, the resulting dissipativity gap was identified
analytically, and a terminal ingredient (a terminal region, a terminal cost,
a terminal equality, or a stage-cost regularisation) was selected and
motivated against the operation's specific mission requirement. The three
instances were then drawn together into a single cross-scenario table
relating mission requirement, MPC ingredient, expected theoretical property,
and residual limitation. This is the first concrete step toward the
traceable requirements-to-design mapping motivated in
Section~\ref{sec:intro}.\\
The analysis rests on a number of simplifying assumptions that should be kept
in mind when interpreting the results. The relative dynamics are linearised
CWH, valid only for near-circular reference orbits and small relative
separations, and neglect $J_2$, drag, and solar radiation pressure. The
chaser is modelled as a point mass with an idealised, continuously
throttleable low-thrust actuator; no attitude dynamics, 6-DOF pose tracking,
or actuator discretisation are represented, and the proximity-manoeuvring
phase (FKP $\to$ FAP) is excluded entirely (Section~\ref{sec:mission}). Full,
noise-free state feedback is assumed at every sampling instant, so no
navigation or model uncertainty enters the design, and the quadratic
control-effort cost $\|u\|^2$ is used throughout as a convex surrogate for
the true $\Delta v$ objective $\|u\|_1$. Most importantly, the stability and
performance properties attributed to each terminal ingredient in
Table~\ref{tab:cross} are drawn directly from the corresponding results in
the EMPC literature rather than re-derived here for the specific CWH system
and constraint sets used in this report; no certificate-by-certificate proof
is provided, and the reported ``guarantees'' should be read as the
applicable condition under which the cited theorem holds, not as a
self-contained derivation.\\
Several of these simplifications point directly to open work. Foremost among
them, a rigorous, system-specific derivation of the stability and
performance guarantees sketched in Table~\ref{tab:cross}, verifying the
exact hypotheses of the cited theorems against the CWH dynamics, constraint
sets, and terminal ingredients used in each operation, is left as future
work. Extending the framework to the proximity-manoeuvring and capture
phases would require 6-DOF pose-tracking MPC. Treating navigation and model
uncertainty would move the guarantees from nominal to robust or stochastic
MPC. And a mission-wide storage function would be needed to reason formally
about stability across the full switched sequence rather than within each
phase in isolation. The minimum feasible horizon for the Operation~3
terminal equality ($N=24$, Section~\ref{sec:op3}) was established by a
simulation sweep rather than a closed-form bound for the coupled,
constrained dynamics; deriving such a bound analytically is left as future
work. The passive-safety constraint introduced in
Section~\ref{subsec:conops} is not yet evaluated as a dedicated mission
requirement, and remains an open item for the next iteration of this
report. Finally, all controllers were validated only in simulation with
\texttt{quadprog}.

\section*{Author Contributions}
\addcontentsline{toc}{section}{Author Contributions}

\textbf{Nils Maier:}
Conceptualization,
Methodology,
Software,
Validation,
Formal analysis,
Investigation,
Data curation,
Visualization,
Writing~-- Original Draft.
\textbf{Torbjørn Cunis:}
Conceptualization,
Resources,
Writing~-- Review \& Editing,
Supervision,
Project administration.
\textbf{Tam W.\ Nguyen:}
Methodology,
Resources,
Writing~-- Review \& Editing,
Supervision,
Project administration.

\section*{Acknowledgements}
\addcontentsline{toc}{section}{Acknowledgements}

This report documents work carried out during a research exchange between the
University of Stuttgart and Kyoto University, as a collaboration between the
Institute of Flight Mechanics and Controls (iFR) at the University of Stuttgart
and the Department of Electrical Engineering at Kyoto University.

The first author thanks the Institute for Systems Theory and Automatic Control (IST) at the University of Stuttgart, and in particular Prof.\ Frank Allgöwer and Fabian
Jakob, for supporting his stay in Japan financially through the Vector foreign-study scholarship.

\newpage

\bibliographystyle{ieeetr}
\bibliography{bibliography}

\begin{thebibliography}{10}

\bibitem{Fehse_2003}
W.~Fehse, {\em Automated Rendezvous and Docking of Spacecraft}.
\newblock Cambridge Aerospace Series, Cambridge University Press, 2003.

\bibitem{Petersen2023}
C.~Petersen, R.~Caverly, S.~Phillips, and A.~Weiss, ``Safe and constrained rendezvous, proximity operations, and docking,'' in {\em 2023 American Control Conference (ACC)}, pp.~3645--3661, 05 2023.

\bibitem{MAYNE2000789}
D.~Mayne, J.~Rawlings, C.~Rao, and P.~Scokaert, ``Constrained model predictive control: Stability and optimality,'' {\em Automatica}, vol.~36, no.~6, pp.~789--814, 2000.

\bibitem{Rawlings2017}
J.~B. Rawlings, D.~Q. Mayne, and M.~M. Diehl, {\em Model Predictive Control: Theory, Computation, and Design}.
\newblock Madison, WI: Nob Hill Publishing, 2nd~ed., 2017.

\bibitem{Hartley2015}
E.~N. Hartley, ``A tutorial on model predictive control for spacecraft rendezvous,'' in {\em 2015 European Control Conference (ECC)}, pp.~1355--1361, 2015.

\bibitem{Weiss2015}
A.~Weiss, M.~Baldwin, R.~S. Erwin, and I.~Kolmanovsky, ``Model predictive control for spacecraft rendezvous and docking: Strategies for handling constraints and case studies,'' {\em IEEE Transactions on Control Systems Technology}, vol.~23, no.~4, pp.~1638--1647, 2015.

\bibitem{DiCairano2012}
S.~Di~Cairano, H.~Park, and I.~Kolmanovsky, ``Model predictive control approach for guidance of spacecraft rendezvous and proximity maneuvering,'' {\em International Journal of Robust and Nonlinear Control}, vol.~22, pp.~1398--1427, 08 2012.

\bibitem{DiCairano2018}
S.~Di~Cairano and I.~V. Kolmanovsky, ``Real-time optimization and model predictive control for aerospace and automotive applications,'' in {\em 2018 Annual American Control Conference (ACC)}, pp.~2392--2409, 2018.

\bibitem{Diehl2011}
M.~Diehl, R.~Amrit, and J.~B. Rawlings, ``A lyapunov function for economic optimizing model predictive control,'' {\em IEEE Transactions on Automatic Control}, vol.~56, no.~3, pp.~703--707, 2011.

\bibitem{Angeli2012}
D.~Angeli, R.~Amrit, and J.~B. Rawlings, ``On average performance and stability of economic model predictive control,'' {\em IEEE Transactions on Automatic Control}, vol.~57, no.~7, pp.~1615--1626, 2012.

\bibitem{Ellis2014}
M.~Ellis, H.~Durand, and P.~Christofides, ``A tutorial review of economic model predictive control methods,'' {\em Journal of Process Control}, vol.~24, 08 2014.

\bibitem{Faulwasser2018}
T.~Faulwasser, L.~Gr{\"u}ne, and M.~M{\"u}ller, ``Economic nonlinear model predictive control,'' {\em Foundations and Trends{\textregistered} in Systems and Control}, vol.~5, pp.~224--409, Jan. 2018.
\newblock Publisher Copyright: {\textcopyright} 2018 Now Publishers Inc. All rights reserved. Copyright: Copyright 2020 Elsevier B.V., All rights reserved.

\bibitem{Woicke2025}
S.~Woicke, J.~Jipp, C.~Steimle, N.~Pokrupa, and L.~Metrailler, ``{ADRIOS ClearSpace-1: In-Orbit Demonstration of the Removal of a Non-Cooperative Spacecraft},'' in {\em Proceedings of the 9th European Conference on Space Debris}, (Bonn, Germany), ESA Space Debris Office, 2025.

\bibitem{schaub2018analytical}
H.~Schaub and J.~L. Junkins, {\em Analytical Mechanics of Space Systems}.
\newblock Reston, VA: {AIAA} Education Series, 4th~ed., 2018.

\bibitem{Woffinden2017}
D.~Woffinden and D.~Geller, ``Navigating the road to autonomous orbital rendezvous,'' {\em Journal of Spacecraft and Rockets}, vol.~44, pp.~898--909, 06 2007.

\bibitem{Hill1878}
G.~W. Hill, ``Researches in the lunar theory,'' {\em American Journal of Mathematics}, vol.~1, no.~3, pp.~245--260, 1878.

\bibitem{ClohessyWiltshire1960}
W.~H. Clohessy and R.~S. Wiltshire, ``Terminal guidance system for satellite rendezvous,'' {\em Journal of the Aerospace Sciences}, vol.~27, no.~9, pp.~653--658, 1960.

\bibitem{willems1972dissipative}
J.~C. Willems, ``Dissipative dynamical systems, part {I}: General theory,'' {\em Archive for Rational Mechanics and Analysis}, vol.~45, no.~5, pp.~321--351, 1972.

\bibitem{Borrelli_Bemporad_Morari_2017}
F.~Borrelli, A.~Bemporad, and M.~Morari, {\em Predictive Control for Linear and Hybrid Systems}.
\newblock Cambridge University Press, 2017.

\bibitem{MATLABquadprog}
{The MathWorks, Inc.}, {\em {quadprog --- Quadratic Programming}}.
\newblock MATLAB Optimization Toolbox Documentation, Natick, MA, USA.

\bibitem{WANG20086974}
Y.~Wang and S.~Boyd, ``Fast model predictive control using online optimization,'' {\em IFAC Proceedings Volumes}, vol.~41, no.~2, pp.~6974--6979, 2008.
\newblock 17th IFAC World Congress.

\bibitem{Jerez2011}
J.~L. Jerez, E.~C. Kerrigan, and G.~A. Constantinides, ``A condensed and sparse qp formulation for predictive control,'' in {\em 2011 50th IEEE Conference on Decision and Control and European Control Conference}, pp.~5217--5222, 2011.

\bibitem{AMRIT2011178}
R.~Amrit, J.~B. Rawlings, and D.~Angeli, ``Economic optimization using model predictive control with a terminal cost,'' {\em Annual Reviews in Control}, vol.~35, no.~2, pp.~178--186, 2011.

\bibitem{Zanon2017}
M.~Zanon, L.~Grüne, and M.~Diehl, ``Periodic optimal control, dissipativity and mpc,'' {\em IEEE Transactions on Automatic Control}, vol.~62, no.~6, pp.~2943--2949, 2017.

\bibitem{MULLER2016128}
M.~A. Müller and L.~Grüne, ``Economic model predictive control without terminal constraints for optimal periodic behavior,'' {\em Automatica}, vol.~70, pp.~128--139, 2016.

\bibitem{Koehler2018}
J.~Köhler, M.~A. Müller, and F.~Allgöwer, ``On periodic dissipativity notions in economic model predictive control,'' {\em IEEE Control Systems Letters}, vol.~2, no.~3, pp.~501--506, 2018.

\bibitem{GRUNE2013725}
L.~Grüne, ``Economic receding horizon control without terminal constraints,'' {\em Automatica}, vol.~49, no.~3, pp.~725--734, 2013.

\end{thebibliography}

\end{document}